\documentclass[a4paper,10pt]{article}

\title{Monte Carlo Simulation for the Formation and Growth of Low Dimensionality
Phases During Underpotential Deposition of Ag on Au(100) }
\author{M.C. Gim\'{e}nez, M.G. Del P\'{o}polo and E. P. M. Leiva \footnote {Corrresponding 
autor: E. P. M. Leiva. e-mail:eleiva@mail.fcq.unc.edu.ar} }

\begin{document}

\maketitle

\begin{abstract}

Simulation studies are undertaken for the system Ag/Au(100) by means of
grand canonical Monte Carlo applied to a large lattice system. The
interactions are calculated using the embedded atom model. The formation of
adsorbed Ag phases of low dimensionality on Ag(100) is investigated and the
influence of surface defects on the shape of the adsorption isotherms is
studied. The results of the simulations are discussed in the light of
experimental information available from electrochemical measurements.


\end{abstract}

\section{Introduction}

New horizons have opened concerning the understanding of metal
electrodeposition processes in the last decade. This has been achieved to a
large extent due to the use of electrochemical techniques like cyclic
voltammetry along with \emph{in-situ }nanoscopy techniques. Among these,
scanning tunnelling microscopy (STM) has allowed the observation in direct
space and real time of the formation and growth of a new phase during the
electrocrystallization of a metal from an electrolytic solution. In this
context we can mention the so-called underpotential deposition (\textbf{upd}%
), which implies the deposition of a metal $M$ on the surface of an
electrode or substrate $S$ of different nature at potentials more positive
than those predicted from the Nernst equation. 
A recent and extensive discussion on this phenomenon can be found 
in a book of Budevski, Staikov and Lorenz \cite{Lorenz_Libro}, and the 
theoretical aspects have been considered in reviews by one of us 
\cite{Leiva_UPD_1,Leiva_UPD_2}.

One way to quantify \textbf{upd }is through the so-called underpotential
shift $\Delta \phi _{UPD},$ which is related to the difference of the
chemical potential of $M$ adsorbed on $S$ at a coverage degree $\Theta $,
say $\mu (M_\Theta /S)$, and the chemical potential of $M$ in the bulk
phase, say $\mu (M/M)$, through the following equation:

\bigskip{} {\centering 
}

{\ 
\begin{equation}
\Delta \phi _{UPD}=\frac 1{ze_0}[\mu (M/M)-\mu (M_\Theta /S)]  \label{1}
\end{equation}
}

\bigskip{}

{\raggedright where $z$ is the charge of the ion $M$ in the solution and $e_0
$ is the elemental charge. Criteria for the experimental estimation of $%
\Delta \phi _{UPD}$ have been given in the literature by Kolb et al.\cite
{Kolb} and Trasatti \cite{Trasatti}. Concerning the modelling of \textbf{upd}
as a thermodynamic process, different interesting approaches have been taken
by several authors in the 70's} \cite
{Lorenz_Libro,Lorenz_2,Bewick_1,Bewick_2, Blum}. The most widespread idea is
to envisage this phenomenon as related to the formation of different
adsorbed phases, involving in some cases, phases of low dimensionality that
differ in their properties and can be detected as separate entities by a
given experimental techniques, i.e. peaks in a voltammogram. The
interpretation of {\textbf{upd}} as a phase transition has been employed by
Blum et al \cite{Blum} to develop a model for the underpotential
deposition of Cu on Ag(111) in the presence of sulfate anions and has also
been recently brought to a more general context by Lorenz and coworkers \cite
{SGarcia}. According to these ideas, the dimensionality of the phases
involved in the \textbf{upd} phenomena is between 0 and 2. The 0D phases
correspond to the adsorption of isolated metal atoms, the adsorption on kink
sites, the adsorption on vacancies of the first lattice plane of the
substrate or on some other type of point defects. The 1D phases are related
to the adsorption at the border of steps or monoatomic terraces and the 2D
phases are related to the formation of monolayers and submonolayers of
adsorbate building compressed, (1x1) or even expanded structures.

Figure 1 illustrates the most common situations experimentally observed,
which will also be considered in the present simulations studies.

It can be observed that relative abundance of 0D and 1D phases is determined
by the presence of defects on the electrode surface. Its existence has been
related to the existence of small shoulders in the voltammograms, as
postulated by Siegentaler et al. \cite{Sigentaler_1} and Lorenz and coworkers 
\cite{Lorenz_3}. In the case of the 2D phases that are formed on the flat
terraces or defect-free surfaces there is clear experimental evidence for
the existence of compressed \cite{Toney}, compact and expanded structures 
\cite{Lorenz_Libro,SGarcia}. The formation and growth of a compact, defect-free
bi-dimensional structure can be described as a first-order phase transition
that takes place at the electrode/solution interface and its occurrence
should be characterized by a discontinuity in the adsorption isotherm, with
a concomitant very sharp peak (more properly speaking a Dirac delta
function) in the voltammetric profiles \cite{Blum}. Concerning this point, it
is remarkable that although \textbf{upd} has been often related to the
presence of first-order phase transitions, the real existence of these
phenomena in experimental systems is still not completely free of
controversy, since discontinuities in the experimental isotherms are usually
not obtained when specific anion adsorption is avoided \cite{Lorenz_Libro}.
We arrive thus at the somewhat paradoxical situation where experimentalists
usually perform their experiments under conditions where the clean
adsorption process of metal adsorption is actually coupled to the even more
complex process of anion coadsorption. Higher order phase transitions should
be characterized by the presence of rather wide and flat voltammetric peaks 
\cite{Blum}, though there is no experimental evidence for the occurrence of
this type of phenomena in electrochemical systems \cite{Lorenz_Libro}. In
this respect, computer simulations could also deliver enlightening
information. Furthermore, computer simulations play a fundamental role for
the comprehension of surface phenomena \cite{Nicholson}. Many approaches can
be taken for the study of surface-related problems. Among others, we can
mention molecular dynamics simulations, Monte Carlo simulations based on
continuum Hamiltonians (off-lattice) and techniques in which the surface is
modelled through a lattice Hamiltonian (on-lattice simulations) \cite
{Ercolessi_1}.

The present work deals with the computer simulation of Ag deposition of
Au(100) by means of the Grand Canonical Monte Carlo method and the
interatomic potential given by the embedded atom method (EAM), which has
been previously employed by some of us to study electrochemical systems \cite
{Cu/Ag}. We developed a simulation scheme that combines the simplicity of
the lattice model of the surface with the many body properties of the
metallic binding. Adsorption isotherms under different temperature
conditions were obtained in order to analyze the existence of a first-order
phase transition, and the influence of different surface defects on the
behaviour of the isotherms is also investigated.

\section{Some experimental facts}

Cyclic voltammetry and STM \emph{in-situ }experiments performed on the
systems Ag/Au(hkl) have shown the presence of different adsorbed phases,
depending on the applied electrode potential \cite{SGarcia, Ag/Au(100),
Ag/Au_Nuevo, Hara}. In the particular case of Ag adsorption on Au(100), the
experiments have revealed the existence of an expanded, quasi hexagonal 
$c(\sqrt{2}\times 5\sqrt{2})R45^{\circ }$Ag phase at high overpotentials. The
existence of this expanded structures can hardly be explained in terms of
energetic considerations, so that the entropic contributions to the free
energy of the system should be considered. Furthermore, the presence of
specifically adsorbed anions may in turn play an important role in
stabilizing this type of entities \cite{upd-eam}, so we shall leave the
consideration of this type of structures for future work. On the other hand,
at intermediate underpotentials the decoration of monoatomic steps
corresponding to the formation of a one-dimensional phase was observed,
while at underpotentials close to Ag bulk deposition, two dimensional
islands were built on terraces and defect-free zones of the substrate. STM
measurements with lateral atomic resolution showed that these islands
present a quadratic symmetry, suggesting that they may be considered as a $%
(1\times 1)$ structure commensurate with the Au$(100)$ substrate. Finally,
it must be added that long time polarizations showed some indication for the
beginning of a surface alloying process whose mechanism has so far not been
explained.

\section{Model and simulation method}

\subsection{Lattice model}

Lattice models for computer simulations are of widespread use in studies of
nucleation and growth, because they allow dealing with a large number of
particles at a relatively low computational cost. In principle, it must be
kept in mind that continuum Hamiltonians should be much more realistic in
those cases where epitaxial growth of an adsorbate leads to incommensurate
adsorbed phases \cite{Pb/Ag} or to adsorbates with large coincidence cells.
On the other hand, the use of fixed rigid lattices restricts enormously the
number of possible configurations for the adsorbate and its use may be
justified on the basis of experimental evidence or continuum computer
simulations that predict a proper fixed lattice geometry. In the present
case, we have strong evidence from this latter type of simulations within
the canonical Monte Carlo method \cite{Portugal} that indicates that at
least one of the phases present during Ag underpotential deposition on
Au(100) possesses a pseudomorphic structure. In fact, our continuum MC
simulations showed that a Ag monolayer adsorbed on Au(100) spontaneously
acquired a $(1\times 1)$ coincidence cell in agreement with the experimental
finding at low underpotentials already mentioned in section 2. For this
reason, we shall employ here a lattice model to represent the square (100)
surface lattice in a Grand Canonical Monte Carlo simulation. Besides Ag
adatoms, Au adatoms are allowed on the surface at different coverage
degrees, building a variety of surface structures with the purpose of
emulating some of the most common surface defects. Linear sweeps of the
chemical potential are performed to obtain the adsorption isotherms and
study the influence of the surface defects on their shapes. First of all, we
neglect the effect of the presence of solvent molecules. This approximation
should not be critical as long as the partial charge on the adatoms is
small, thus minimizing the ion-dipole interactions. Second, we also neglect
all kinds of anion effects that may coadsorb during the metal deposition
process. In the case of other systems, the adatom-anion interactions have
been shown to be very important \cite{Aniones}, playing a decisive role in
determining the energetics of the system. This has been recently analyzed by
some of us in thermodynamic terms \cite{upd-eam}. In the case of the present
system, however, thin layer twin electrode experiments gave no indication
for any change in the extent of anion adsorption upon building the adsorbed
monolayer.

Square lattices of different sizes with periodical boundary conditions are
used in the present work to represent the surface of the electrode. Each
lattice node represents an adsorption site for a Ag or a Au atom. The former
may adsorb, desorb or jump between neighboring sites, while the latter may
only move on the surface like the Ag atoms do. In this way, our model
corresponds to an open system for one of its components, that is, Ag. This
has physical correspondence with the setup of the electrochemical
experiment, where only the metal in equilibrium with its cations in solution
may dissolve or be deposited in the potential range considered.

Concerning the Au adatoms, some considerations must be made regarding the
existence of surface defects. These atoms may in principle move freely on
the surface and minimize the free energy of the system by a number of
mechanisms. For example, isolated Au atoms may heal defects through their
incorporation to a defective cluster, or small Au islands may dissolve to
join large ones, like shown in previous simulation work by Stimming and
Schmickler \cite{S&S}. However, as long as these processes are slow enough 
as compared with the processes related to the Ag adatoms (and this is found 
to be the case), the whole process can be envisaged as the adsorption/desorption
behaviour of Ag on a Au(100) surface with defects. Thus, different Au
structures can be imposed as initial conditions for each simulation. In the
present case, they range from islands of different sizes and shapes obtained
by means of simulated annealing techniques, up to completely random
distributions. All this was undertaken in order to emulate some of the
defects that can be found on a real single crystal surface, like kink sites,
vacancies, isolated substrate atoms, steps, etc.

\subsection{Interatomic potential}

A very important feature to be taken into account when comparing the results
of a simulation with experiment is the quality of the interatomic potentials
used to perform the simulations. Several methods have been developed to
calculate the total energy of a many-particles metallic system, with a
computational effort comparable to that of a pair potential \cite{Carlsson}.
Among these models, denominated broken-bonds or bond-cutting methods it is
worth mentioning the embedded atom method(EAM) \cite{Daw-Baskes}, the N-body
potentials of Finnis and Sinclair \cite{Finnis}, the second-moment
approximation or Tight-binding (TB) \cite{Carlsson} and the glue model (GM) 
\cite{Ercolessi_2}. In this work we use the embedded atom method\cite
{Daw-Baskes} because it is able to reproduce important characteristics of
the metallic binding that cannot be obtained using simple pair potentials.

The EAM considers that the total energy $U_{tot}$ of an arrangement of $N$
particles may be calculated as the sum of energies $U_i$ corresponding to
individual particles 
\begin{equation}
U_{tot}=\sum_{i=1}^NU_i  \label{2}
\end{equation}
where $U_i$ is given by 
\begin{equation}
U_i=F_i(\rho _{h,i})+\frac 12\sum_{j\neq i}V_{ij}(r_{ij})  \label{3}
\end{equation}
$F_i$ is denominated embedding function and represents the energy necessary
to embed atom $i$ in the electronic density $\rho _{h,i}$ at the site at
which this atom is located. $\rho _{h,i}$ is calculated as the superposition
of the individual electronic densities $\rho _i(r_{ij}):$%
\begin{equation}
\rho _{h,i}=\sum_{j\neq i}\rho _i(r_{ij})  \label{4}
\end{equation}
Thus, the attractive contribution to the EAM potential is given by the
embedding energy, which accounts for many-body effects. On the other hand,
the repulsion between ion cores is represented through a pair potential $%
V_{ij}(r_{ij})$, which only depends on the distance between the cores $%
r_{ij} $: 
\begin{equation}
V_{ij}=\frac{Z_i(r_{ij})Z_j(r_{ij})}{r_{ij}}  \label{5}
\end{equation}
$Z_i(r_{ij})$ may be envisaged as a sort of effective charge, dependent on
the nature of the particle $i$. The EAM has been parametrized to fit
experimental data like elastic constants, dissolution enthalpies of binary
alloys, bulk lattice constants and sublimation heaths \cite{Daw-Baskes}. Pair
functionals have been widely used for surface diffusion studies and
adsorption of metals on metallic surfaces \cite{Cu/Ag, Votter, Haftel}.

\subsection{Grand Canonical Monte Carlo}

One of the most appealing characteristics of Grand Canonical Monte Carlo $%
(\mu VT/MC)$ is that the chemical potential $\mu $ is one of the independent
variables, as it is the case in many experimental situations. In the
electrochemical case, we can mention low-sweep rate voltammetry, where the
electrode potential can be used to control the chemical potential of species
at the metal/solution interface. This technique offers a straightforward way
of obtaining the adsorption isotherms provided the sweep rate is low enough
to ensure equilibrium for the particular system considered. Another
interesting aspect of $\mu VT/MC$ is that a better and faster equilibration
of the system is possible due to local density changes originated by
fluctuations in the number of particles.

Our 2D system is characterized by a square lattice with $M$ adsorption
sites. We shall label each adsorption site with a number $0$, $1$ or $2$,
depending on whether it is empty or occupied by a gold or a silver atom
respectively. Thus, the configuration of the system will be defined through
the components $n_1$, $n_2$,$\cdots $,$n_M$ of an $M$-dimensional vector $%
\overrightarrow{n}$ and its partition function may be written as:

{\raggedright 
\begin{equation}
\Xi (\mu _1,\mu _2,M,T)=\sum_{\overrightarrow{n}}\exp \beta (\mu
_1\sum_{i=0}^M\delta _{1,n_i}+\mu _2\sum_{i=0}^M\delta _{2,n_i})\exp (-\beta
U(\overrightarrow{n}))  \label{6}
\end{equation}
where $\mu _1$ and $\mu _2$ are the chemical potentials of gold and silver
at the interface, $\delta _{1,n_i}$ and $\delta _{2,n_i}$ are Kronecker
delta functions acting on the components of the vector $\overrightarrow{n}$
and $U(\overrightarrow{n})$ is the energy of the system in state $%
\overrightarrow{n}$. The probability of obtaining a given configuration will
be given by the corresponding Boltzmann factor:}

\begin{equation}
P_{\overrightarrow{n}}(\mu _1,\mu _2,M,T)=\frac{\exp \beta (\mu
_1\sum_{i=0}^M\delta _{1,n_i}+\mu _2\sum_{i=0}^M\delta _{2,n_i})\exp (-\beta
U(\overrightarrow{n}))}{\Xi (\mu _1,\mu _2,M,T)}  \label{7}
\end{equation}

Following the procedure proposed by Metropolis and coworkers \cite{Allen},
the acceptance probability for a transition from state $\overrightarrow{n}$
to $\overrightarrow{n}^{\prime }$ is defined as:

\begin{equation}  \label{8}
W_{\overrightarrow{n}\rightarrow \overrightarrow{n}^{\prime}}=\min (1,\frac{%
P_{\overrightarrow{n}^{\prime}}}{P_{\overrightarrow{n}}})
\end{equation}

{\raggedright so that detailed balance is granted.}

In our $\mu VT\ /\ MC$ simulation we shall allow for three types of events :

\begin{enumerate}
\item  Adsorption of a Ag atom on a lattice site selected randomly.
According to eqns.\ref{7} and \ref{8}, the acceptance probability 
$W_{\overrightarrow{n}\rightarrow \overrightarrow{n}^{\prime }}$ for this event
is: 

\begin{equation}
W_{\overrightarrow{n}\rightarrow \overrightarrow{n}^{\prime }}=\min (1,\exp
(\beta (\mu _1-\Delta U)))  \label{9}
\end{equation}

where $\Delta U=U(\overrightarrow{n}^{\prime })-U(\overrightarrow{n})$.

\item  Desorption of a Ag atom selected randomly. This probability is 
\begin{equation}
W_{\overrightarrow{n}\rightarrow \overrightarrow{n}^{\prime }}=\min (1,\exp
(\beta (-\mu _1-\Delta U)))  \label{10}
\end{equation}

\item  Motion of a Ag or a Au atom from the lattice site where it is
adsorbed to one of its four nearest neighbors. The latter is selected
randomly and the acceptance probability for this event is equivalent to that
employed in the canonical ensemble, since the number of particles remains
constant. We have: 
\begin{equation}
W_{\overrightarrow{n}\rightarrow \overrightarrow{n}^{\prime }}=\min (1,\exp
(-\beta \Delta U))  \label{11}
\end{equation}
\end{enumerate}

Even when in a grand canonical simulation, events of type 3 (motion or
diffusion of lattice particles) are not strictly necessary, their presence
is justified by the fact that a smaller number of Monte Carlo steps(MCS)
must be employed for the equilibration of the system\cite{Allen}. The
calculated equilibrium thermodynamic properties do remain unaltered by this
choice.

In order to satisfy detailed balance, the \emph{a priori }probability for
the creation of a Ag atom, $\alpha _c$, must be equal to the probability of
its anhilation $\alpha _d$ and independent of its motion probability $\alpha
_m.$

The algorithm devised for an elementary step of the simulation is the
following:

\begin{enumerate}
\item  Selection of the elementary process(adsorption, desorption of
diffusion) to occur. This is made by means of the generation of a random
number $\epsilon $ uniformly distributed in the interval $[0,1]$.

\item  If $0\leq \epsilon \leq (\alpha _c+\alpha _d)$, a random site is
chosen on the lattice.

\begin{itemize}
\item  If the occupation number of the site is $0$, an attempt is made to
create a Ag atom at that site, according to the acceptance given by eqn. \ref
{9}.

\item  If the occupation number of the site is $2$, an attempt is made to
destroy the atom adsorbed at that site, according to the acceptance given by
eqn. \ref{10}.
\end{itemize}

\item  If $(\alpha _c+\alpha _d)\leq \epsilon \leq 1$ , an attempt is made
to move each of the particles of the system according to the following
algorithm:

\begin{itemize}
\item  One of the nearest sites to the particle is selected. If the
occupation number is $0$, a movement to the new position is attempted
according to eqn. \ref{11}. Otherwise, the next particle in the list is
considered.
\end{itemize}

\item  Back to point 1, thus completing one MCS.
\end{enumerate}

Within this procedure, the relevant thermodynamic properties are then
obtained as average values of instantaneous magnitudes stored along a
simulation run. For example, the average coverage degree of the silver atoms 
$\left\langle \Theta \right\rangle _{Ag}$at a given chemical potential $\mu
_1$ after a simulation with $MCS_e$ equilibration steps and a total number
of $MCS$ steps will be given by:

\begin{equation}
\left\langle \Theta (\mu _1)\right\rangle _{Ag}=\frac
1{MCS-MCS_e}\sum_{i=MCS_e}^{MCS}\Theta (\mu _1)_{Ag,i}  \label{12}
\end{equation}

where the instantaneous value of silver coverage degree{, $\Theta (\mu
_1)_{Ag,i}$ , is defined from the number of Ag atoms $N_{Ag,i}$ and the
number of Au atoms $N_{Au}$ present at the interface at the time step $i$ }

\begin{equation}
\Theta (\mu _1)_{Ag,i}=\frac{N_{Ag,i}}{M-N_{Au}}  \label{13}
\end{equation}

Since our main goal is to obtain adsorption isotherms for several Ag
coverage degrees and different surface structures of Au, we chose a
simulation method that does not deliver information on the kinetics of
nucleation and growth. In this respect, the Kinetic Monte Carlo(KMC)\cite
{Rickvold,Schefler} technique should be applied and will be the subject of
future work.

\subsubsection{Algorithm employed for the calculation of energy differences}

In a MC simulation the determining step concerning the speed of the
calculation is the evaluation of the total energy of the system 
$U(\overrightarrow{n})$, which is determined by the type of interatomic
potential employed.

According to the algorithm described in the previous section, any new state
of the system is accepted with a probability that involves the Boltzmann
factor of the energy difference between the initial and the final state,
that is $\Delta U=U(\overrightarrow{n}^{\prime })-U(\overrightarrow{n}).$
One of the main advantages of the lattice model is its simplicity, since it
fixes the distances between the adsorption nodes, thus reducing the energy
values that the system can take to a discrete set. Furthermore, the
potentials used are short ranged, so that a very important simplifying
assumption can be made for obtaining $\Delta U.$ The point is to consider
the adsorption(desorption) of a particle at a node immersed in a certain
environment surrounding it, as shown in Figure 2 . The adsorption site for
the particle is located in the central box, and the calculation of the
interactions is limited to a circle of radius $R.$ Then, the adsorption
energy for all the possible configurations of the environment of the central
atom can be calculated previous to the simulation. The $S$ adsorption sites
within the circle are labelled arbitrarily with numbers between $1$ and $S$
(Figure 2) and their labels along with the occupation number may be employed
to identify the configuration in terms of a number $I$ expressed on the
basis of $3$. Generally speaking, for a configuration characterized by the
occupation numbers $n_1,\,n_2,\,\ldots \,n_S$ , the corresponding value of $I
$ is calculated as $I=\sum_{i=1}^{S-1}n_i\ 3^{S-i-1}$ . For example, for an
environment with 8 neighbors $(S=9)$ such as that shown in Figure 2a, the
configuration $n_1=2,\,n_2=2,\,n_3=1,\,n_4=1,\,n_5=0,\,n_6=0,\,n_7=0,\,n_8=0$
is indexed according to our convention with $I=6156$. With this method all
the adsorption/desorption energies of Ag and Au on Au(100) are tabulated, so
that during the MC simulation the most expensive numerical operations are
reduced to the reconstruction of the number $I$ that characterizes the
configuration surrounding the particle on the adsorption node.
Computationally speaking, $I$ is nothing but the index of the array in which
the energy is stored.

In the present work, the adsorption/desorption energies were calculated for
environments with $R$ equal to the distance between second and third nearest
neighbors, as shown in Figure 2 , using as a substrate a five-atomic planes
thick gold slab. The results for the energy are shown as a histogram in
Figure 3 , where the frequency of appearance of a given energy is shown as a
function of the adsorption energy. This is a very important information,
since this figure contains the distribution of energy changes that an atom
may undergo during the simulation. A more detailed discussion on the
information that can be extracted from the histograms will be given in
section 4.1.

We must finally mention a further assumption in our calculations. In all the
structures generated the adsorbates were located at a vertical distance from
the first substrate plane equivalent to the distance between neighboring
Au(100) lattice planes. This should not be a very important approximation,
since the difference between the Au and Ag lattice constants amounts only
0.01 \AA .

\subsubsection{Simulated annealing}

Simulated annealing techniques have often been used to obtain minimal energy
structures or to solve ergodicity problems. A suitable way to implement them
is through the canonical Monte Carlo method at different temperatures. The
simulation is started at a very high initial temperature $T_o$, of the order
of 10$^5$ K, and the system is later cooled down following a logarithmic law:

\begin{equation}
T_f=T_oK^{N_{cycles}}  \label{14}
\end{equation}
where T$_f$ is the final temperature, $N_{cycles}$ is the number of cooling
steps and $K$ is a positive constant lower than one. A few hundreds of
thousands of MCS are run at each temperature in order to allow an extensive
exploration of the configuration space and the simulation stops when $T_f$
is reached. Although experimental annealing treatments (like flame
annealing) are used to get clean and structurally well defined surfaces, in
the present case we have used the computational homologue to obtain the more
stable surface defects of Au at $300\,K$.
In the case of experiments, this procedure has been sometimes questioned and 
we will see here that the type of structures obtained are indeed very sensitive 
to the cooling rate employed.

\section{Results and discussion}

\subsection{Some considerations on the adsorption/desorption energies of Ag}

The information provided by the histograms shown in Figure 3 is very
valuable since they contain all the possible adsorption/desorption energies
that a Ag atom may undergo during the simulation. For the purpose of the
present analysis, we roughly distinguish between three types of
configurations: those in which the central atom (box 13 in fig. 2b) is
surrounded only by other Ag atoms(or empty sites), those in which the
central atom is only surrounded by Au atoms and those in which the
environment may contain Au or Ag atoms indistinctly. The two former families
involve 4096 configurations each, while the latter consists of 531441
configurations, including the two previous ones. In can be observed that in
the two former cases (continuous and dotted histograms respectively) the
energy distribution is multimodal, they being qualitatively very similar to
each other. However, the histogram corresponding to Ag adsorption in the
presence of Au atoms is flatter and is shifted towards more negative
adsorption energies. This suggests a certain affinity of Ag for a Au
environment. From this point of view, the adsorption of Ag in the presence
of Au surface defects should be favoured with respect to the adsorption on a
perfect surface. This speculation will be corroborated by the MC simulations.

The pulse-plot in Figure 3 corresponds to the distribution of the adsorption
energies in the mixed environment for the 12-neighbors system. The energy
distribution looks still multimodal but considerably smoother than the
previous ones. Thus, in general terms it can be concluded that the presence
of gold atoms surrounding the adsorbed Ag atoms as surface defects will
widen and enrich the energy spectrum of the adsorbate.

\subsection{Ag deposition on a perfect Ag(100) single crystal surface.}

Our initial work was devoted to the deposition of Ag on a defect-free
Au(100) surface, performing $\mu VT\ /\ MC$ simulations at $300\,K.$ The
systems consisted in $(40\times 40)$ and $(100\times 100)$ lattices, using
the adsorption/desorption energy tables calculated as described in Section
3.3.1.

Positive and negative chemical potential sweeps were made in the range
between $-2.7\,eV$ and $-3.2\,eV$, involving simulations with a length of $%
10^6$ to $10^7$ $MCS$ , depending on the system size. In order to quantify
underpotential deposition, the deposition of Ag on Ag(100) was also
simulated. The chemical potential for bulk Ag deposition $\mu _{Ag/Ag}^c$
was determined from these runs.

In all cases the adsorption isotherms show a very abrupt change of the
coverage degree, Fig. 4a, at a characteristic chemical potential $\mu
_{Ag/Au}^c$, that mainly depends on the cutoff radius employed for the
calculation of the energy and to a lower extent on the system size. The $\mu
_{Ag/Au}^c$ values along with other relevant information are reported in
Table 1. The drastic change observed in the adsorption isotherms points
towards the existence of a first-order phase transition in which a
two-dimensional phase (an adsorbed monolayer) is built through a nucleation
and growth process. According to the values of chemical potentials reported
in Table 1, this occurs at lower $\mu $ values than those expected for bulk
Ag deposition $(\mu _{Ag/Ag}^c=-2.865\,\,eV)$, leading to an underpotential
shift $\Delta \phi _{UPD}=\mu _{Ag/Ag}-\mu _{Ag/Au}$ that can be estimated
to be between $\,0.06\,eV$ and $0.11\,eV$ . The isotherm corresponding to Ag
adsorption on Ag(100) is also shown in Figure 4a.

Concerning the chemical potential sweep, it was performed in two different
fashions. In the first, for each new $\mu $ value characterizing the
simulation, an initial value of coverage degree of Ag, $\Theta _{Ag},$ was
set, that was 0, 0.5 or 1. This delivered three types of isotherms that
corresponded to the different initial conditions. In the second type of
simulations, the final configuration at a given $\mu $ was employed as
starting point for the next simulation. The first type of simulations was
employed to study hysteresis effects, that in our system were found to be
important at 300 K (Figure 4b). As shown below, they are related to the
absence of surface defects since it is precisely on these defects where the
formation of the new phase starts. The simulations undertaken with the
second scheme showed essentially the same characteristics as those with the
first.

The effect of temperature was also investigated and it was found that at
high temperatures the sharp behaviour of the $\Theta -\mu $ isotherms
changes to a smooth, langmuirian-like behavior, as can be seen in Figure 5.
From these curves we can infer that at $300K$ the present system is well
below its critical temperature. We can also estimate this critical
temperature using the equation of the 2-D Ising model:

\[
T_c=\frac w{2k\ln \left( \sqrt{2}-1\right) }
\]
where $k$ is the Botzmann constant and $w$ is the interaction energy between
nearest neighbours. We have estimated a value of $w$ by inserting a Ag
adatom in different environments.\ The values obtained were in the range $%
-0.27\ eV<w<-0.18\ eV$ so that the critical temperature is estimated to be
in the range $1200\ K<T_c<1800\ K$. These values should only provide a rough
estimation, since the Ising model does not consider the many body effects
present here.

\subsection{Influence of surface defects on the adsorption isotherms}

A wide variety of defects may exist on a single crystal surface, to an
extent that it depends among other factors on the temperature, the chemical
environment and the method by which it was generated. We therefore performed
an important number of simulations to elucidate the effect of the most
common surface defects on the Ag adsorption isotherm. The starting
configuration for the Au defects on the surface was obtained from a
simulated annealing calculation and a $100\times 100$ lattice and $R=3\,$
were used in all cases. The length of the simulations ranged between $10^6$
and $10^7$ $MCS$, with an adequate number of equilibration steps.

\subsubsection{Generation of surface inhomogeneities by means of simulated
annealing}

It has been demonstrated experimentally that after a flame annealing
treatment the Au(100) surface is reconstructed to give a quasi-hexagonal
structure \cite{Oro_reconstruccion}. This reconstruction is lifted upon a
potential sweep at a potential value that depends on the nature of the
electrolyte involved. Since the hexagonal structure is more compact than the 
$(1\times 1)$ structure, a number of islands are built on the electrode
surface that accommodate the remaining atoms. In the case of the
experimental studies of Ag deposition on Au(100), STM and AFM \emph{in-situ}
images indicate that previous to Ag deposition the substrate consists in
terraces separated by monoatomic steps, while the corresponding images with
lateral atomic resolution show a quadratic structure characteristic for an
unreconstructed surface\cite{SGarcia, Lorenz_3}.

Since the atomic excess expected after the lifting of the reconstruction is
about $13\ \%$ of the atomic density of the Au(100) single crystal surface,
we generated Au islands that cover the single crystal surface with a
coverage degree $\Theta _{Au}\simeq 0.1$. Thus, simulated annealing was
performed with $1.000$ Au atoms distributed on a $100\times 100$ lattice at
an initial temperature $T_0$ of $10.000\ K.$ The system was later cooled
down to a final temperature $T_f$=$300\ K$ following the cooling rate given
by eqn. \ref{14} with $K$= $0.8.$ This produced $16$ cooling steps between
the initial and the final temperature. In order to produce surfaces with
different densities of defects, the number of Monte Carlo steps allowed at
each temperature was varied, obtaining the different final structures shown
in Figure 6 . It can be noticed that when the number of $MCS$ at each
temperature is increased, the islands turn more compact, with well defined
borders. This is equivalent to a low concentration of defects such as kinks
and monoatomic steps. On the other hand, if the cooling is faster, the
number of islands per unit surface increases and their borders turn more
irregular. Thus, the lower the number of $MCS$ allowed at each temperature,
the farther the final state will be from equilibration. Table 2 shows the
density of kink sites $\rho _k$ and the density of sites located at the
border of the monoatomic steps $\rho _s$ at different number of $MCS.$ The $%
\rho _k$ was calculated as $\frac{M_k}M$ where $M_k$ is the number of kink
sites and $M=N^2$ is the total number of lattice sites$(10.000).$
Analogously, $\rho _s$ was calculated as $\frac{M_s}M$ where $M_s$ is the
number of sites located at the border of the monoatomic steps. The
relationship perimeter/area of the islands, calculated from the number of Au
atoms located at the border and the number of atoms located inside the
islands, is also given in the table. All the values refer to the final
configuration of the system.

\subsubsection{Deposition on kink sites}

The first stage in the deposition process is given by the adsorption of Ag
atoms on point defects. This can be envisaged as the formation of a $0$%
-dimensional phase that occurs at chemical potentials more negative than the
chemical potential for the formation of the Ag monolayer. Although the most
commonly observed point defects on the surface of a monocrystalline
electrode at $300\ K$ are kink sites, their density at this temperature may
be affected by different factors, like the presence of a condensed phase in
contact with the metallic surface \cite{Lorenz_Libro}. In the present work
we considered that the initial number of kink sites and the corresponding
roughness of the monoatomic terraces of the substrate only depends on the
cooling rate employed in the simulated annealing process. The initial
structure of the stepped Au surface was set according to the results
obtained in section 4.3.1 and sweeps of chemical potential were undertaken
between $-3.30$ and $-3.02\,eV$ . Figure 7a shows isotherms obtained with $5$
of the $16$ configurations resulting from the simulated annealing process.
The gold structures considered correspond to the final state resulting from
simulations with $20,\ 320,$ $5120,$ $81920$ and $655360\;MCS$ and range
from polyhedron shaped islands with different roughness up to a more or less
random distribution of very small irregular islands. A coverage degree due
to adsorption on kink sites $\Theta _k$ may be defined considering that they
form a sublattice of $M_k$ adsorption nodes at a given stage of the
simulation. According to this, we defined $\Theta _k=\frac{N_{Ag}^k}{M_k}$,
where $N_{Ag}^k$ is the number of Ag atoms adsorbed on kink sites. It can be
clearly observed that none of the isotherms present an abrupt change in the
coverage degree, showing a rather langmuirian shape that indicates that
cooperative effects are absent in this type of adsorption. Another
remarkable feature that is relevant for the discussion below is the fact
that for increasing density of islands and the concomitant increasing
roughness of their borders, the adsorption isotherms are shifted towards
more negative chemical potentials.

It is clear that the kink sites subject of the present analysis are not all
equivalent. Furthermore, many of them are actually not independent since the
distance between them is lower than the cutoff radius for the interactions
and they can even move in the course of the simulation due to the mobility
allowed to the Au atoms. It may thus be interesting to compare the present
isotherms for kink sites with others where the equivalence and independence
of the adsorption nodes is granted \textit{a priori} by considering an ideal
ensemble of identical and independent sites. With this purpose we built such
lattices with the most commonly found kink sites, resulting from our
simulated annealing procedure. The most commonly found types of kink sites
are shown in figure 8 . With the adsorption energies corresponding to each
of these sites we constructed the corresponding Langmuir isotherms for the
coverage degree $\Theta _i^{Lang}\ $as a function of the chemical potential $%
\mu .$ These isotherms were then used to compose what we shall denominate an
''ideal '' adsorption isotherm for Ag adsorption on kink sites $\Theta
_{kink}^{ideal}/\mu $ according to the following. Let us denote with $\rho
_k^1,\,\rho _k^2$ and $\rho _k^3\,$ the densities of each of the types of
kink sites calculated from the simulated annealing results according to: 
\[
\rho _k^i=\frac{M_k^i}{M_k\ }
\]
with $M_k^i$ denoting the number of kink sites of type $i$ and $%
M_k=M_k^1+M_k^2+M_k^3.$ From the three isotherms $\Theta _1^{Lang}/\mu $, $%
\Theta _2^{Lang}/\mu $ and $\Theta _3^{Lang}/\mu $, we defined the ''ideal
'' adsorption isotherms for adsorption on kink sites as: 
\begin{equation}
\Theta _{kink}^{ideal}(\mu )=\rho _k^1\Theta _1^{Lang}+\rho _k^2\Theta
_2^{Lang}+\rho _k^3\Theta _3^{Lang}  \label{15}
\end{equation}

Ideal kink sites isotherms for the five imperfect structures selected are
shown in Fig. 7b. Comparison with Fig. 7a indicates that the more compact
the islands the closer the behaviour of the real isotherms to the ideal
ones. On the other hand, for increasing roughness of the island borders, the
real isotherms shift their inflection point towards lower chemical
potentials. This larger deviation from ideality in the case of the rougher
structures is understandable, since in this case the distance between kink
sites diminishes and they no longer behave as independent adsorption sites.

Table 3 summarizes the chemical potentials at the inflection point ($\mu
_{infl}$) for real and ideal isotherms. Comparison with the value for the
deposition of a Ag monolayer on a perfect surface shows that the presence of
point defects may shift the threshold for underpotential deposition in the
negative direction as much as 0.2 V.

It is also interesting to compare these values with that corresponding to
the filling of Au surface monovacancies with Ag atoms. In this case we
obtained $\mu _{infl}^{vac}=-3.53\ eV$. However, these defects are expected
to have a too low concentration at the working temperatures in
electrochemistry so their relevance may be considered as negligible for
practical purposes.

\subsubsection{Deposition on the border of monoatomic steps}

The second stage of the present electrodeposition process involves the
formation of one-dimensional phases through the decoration of monoatomic
steps. In all cases this process was found in the $\mu $ range between $%
-3.05\,eV$ and $-2.95eV,$ that is, after deposition on point defects and
before the formation of the 2-D phase. The corresponding simulations were
run with the five systems selected above, that is, a $100\times 100$ lattice
partially covered with Au islands and the chemical potential was run between 
$-3.32\,eV$ and $-2.90\,eV$. Figure 9 shows the adsorption isotherms for the
surface configurations selected. Analogously to the case of point defects,
we defined here a coverage degree related to border sites, $\Theta _{step},$
referred to the total number of adsorption sites located at the border of a
monoatomic step in a given configuration of the simulation. If these
isotherms are compared with those of figure 7a, it can be noticed that the
roughness of the Au islands plays the opposite role concerning the stability
of the Ag deposit. That is, the increasing roughness of the islands shifts
the isotherms towards more positive chemical potentials. This fact may be in
principle explained in terms of the considerations made in section 4.1 about
the influence of Au atoms on the energy spectrum of Ag adsorption. Rougher
islands provide adsorption sites with a lower coordination of the Ag border
adatom with the Au atoms of the island, so that the adsorption energy of the
adsorbates is in the average shifted towards more positive values. However,
this is a rather simplified explanation and a more detailed study in terms
of the surface configurations should be undertaken.

Finally, it is remarkable the fact that as the compactness and size of the
islands increase, the isotherms become steeper, indicating some sort of
cooperative effect in the formation of the 1-D phase.

\subsubsection{Formation of Ag monolayers in the presence of surface defects}

The formation of the 2-D phase in the presence of Au islands was finally
investigated in connection with the other processes described above.
Chemical potential sweeps were undertaken in the range $[-3.30\,eV<\mu <$ $%
-2.90\,eV],$ with the observation of adsorption of Ag atoms on kink sites $%
[-3.30\ eV<\mu <-3.02\ eV]$, decoration of monoatomic steps,$[-3.05\ eV<\mu
<-2.95\ eV]$ and the final formation of the monolayer$[$ $-2.95\,\ eV<\mu <$ 
$-2.90\ \,eV]$. Figure 10 shows the final state at different $\mu $ for one
of the five systems considered previously($MCS=5120$).

If we compare the isotherms for this system(Fig. 11) with those
corresponding to adsorption on perfect surfaces(Fig 4a), it is clear that
the presence of surface defects introduces additional stages in the
deposition process due to the formation of 0-D and 1-D phases. It is
remarkable the smoothing of the isotherms in the neighborhood of the
deposition potential of the monolayer, this being enhanced for rougher
substrates. It is also observed that as the number of islands per unit
surface increases, the chemical potential for the formation of the monolayer
is shifted negatively. In all cases the formation of the 2-D phase is built
in a coalescence regime starting from the Au islands. This can be noted in
the snapshots of Figure 10.

All the deposition stages described occur at chemical potentials negative
with respect to Ag bulk deposition ($\mu _{Ag/Ag}^c=-2.865\ eV)$, thus
within the underpotential deposition range. The underpotential shifts $%
\Delta \phi _{UPD}$ for the different processes thus range between $0.4\ eV$
and $0.03\ eV$.

\section{Conclusions}

The deposition of Ag on Au(100) was studied in the absence and in the
presence of surface defects. These were generated by means of a simulated
annealing procedure using different numbers of Monte Carlos steps $(MCS)$
per cooling cycle. It was found that a lower number of $(MCS)$ generates a
higher number of Au islands per unit surface, which are characterized by a
higher perimeter/area rate. On the other hand, a lower cooling rate
generates more compact, bigger Au islands.

Ag deposition always takes place at chemical potentials lower than bulk
deposition, thus leading to a positive underpotential shift. This is also
supported by first-principles calculations performed by one of us \cite
{UPD_AG/AU}. The adsorption isotherms obtained indicate that Ag adsorption
should occur in three well defined stages: first, adsorption on kink sites,
second, adsorption at the border of monoatomic steps and finally formation
of the monolayer or 2-D phase. In all cases the presence of inhomogeneities
shifts the adsorption isotherms towards more negative chemical potentials
thus increasing the underpotential shift with respect to that predicted for
a perfect surface. Furthermore, the isotherms are smoothed in the
neighbourhood of the critical chemical potential, this effect being more
pronounced for stronger disordered surfaces.

The results of our simulations support to a large extent the experimental
results obtained for this system, fundamentally concerning the existence of
a positive underpotential shift and the strong affinity of Ag for Au
evidenced in the trend to build surface alloys. More detailed experimental
studies concerning the influence of surface defects on the voltammograms and
adsorption isotherms would be very helpful to check the present calculations
and to improve the modelling of the system Ag/Au(100). One point to be
considered is that of finite size effects of the simulation cell, especially
in the case of adsorption on kink sites for imperfect surfaces obtained at a
low cooling rate.

\section{Acknowledgments}

Part of {\normalsize the present calculations were performed on a Digital
workstation donated by the Alexander von Humboldt Stiftung, Germany. }%
Fellowships from the Consejo de Investigaciones Cient\'{\i}ficas y
T\'{e}cnicas de la Provincia de C\'{o}rdoba(C.G. and M.D. P.), financial
support from the Consejo Nacional de Investigaciones Cient\'{\i}ficas y
T\'{e}cnicas, the Secretar\'{\i}a de Ciencia y T\'{e}cnica de la Universidad
Nacional de C\'{o}rdoba, the Consejo de Investigaciones Cient\'{\i}ficas y
T\'{e}cnicas de la Provincia de C\'{o}rdoba and language assistance by
Pompeya Falc\'{o}n are gratefully acknowledged.\newpage\ 

\section{Tables}

\begin{center}
\begin{tabular}{|c|c|c|}
\hline
$R\ \ \backslash \ N\times N$ & $40\times 40$ & $100\times 100$ \\ 
\hline\hline
$2$ & -2.975 & -2.970 \\ \hline
$3$ & -2.925 & -2.936 \\ \hline
\end{tabular}
\end{center}

\textbf{Table 1}: Threshold chemical potentials for Ag deposition on Au(100)
in absence of surface defects. Different lattice sizes $(N\times N)$, and
cutoff radii $(R)$ for the interactions were employed.

\begin{center}
{\centering
\begin{tabular}{|c|c|c|c|c|c|c|c|c|}
\hline
$MCS $ & 20 & 40 & 80 & 160 & 320 & 640 & 1280 & 2560 \\ \hline\hline
$\rho _{k} $ & 0.0261 & 0.0235 & 0.0221 & 0.0197 & 0.0163 & 0.0157 & 0.0132
& 0.0117 \\ \hline
$\rho _{s} $ & 0.108 & 0.1044 & 0.0991 & 0.0919 & 0.0877 & 0.0716 & 0.0694 & 
0.0560 \\ \hline
$\frac{P}{A} $ & 82.33 & 32.33 & 20.27 & 12.88 & 8.25 & 4.12 & 3.29 & 1.96
\\ \hline
\end{tabular}
}

\medskip{} {\centering
\begin{tabular}{|c|c|c|c|c|c|c|c|c|}
\hline
$MCS $ & 5120 & 10240 & 20480 & 40960 & 81920 & 163840 & 327680 & 655360 \\ 
\hline\hline
$\rho _{k} $ & 0.0105 & 0.0081 & 0.0068 & 0.0064 & 0.0050 & 0.0029 & 0.0027
& 0.0030 \\ \hline
$\rho _{s} $ & 0.0432 & 0.0392 & 0.0353 & 0.0260 & 0.0184 & 0.0172 & 0.0145
& 0.0142 \\ \hline
$\frac{P}{A} $ & 1.26 & 0.94 & 0.77 & 0.56 & 0.35 & 0.27 & 0.23 & 0.24 \\ 
\hline
\end{tabular}
}
\end{center}

\medskip{} \textbf{Table 2}:Density of kink sites $\rho _k$, density of
border sites at the monoatomic steps $\rho _s$ and perimeter to area
relationship $\frac PA$ of the Au islands generated by simulated annealing. $%
MCS$ is the total number of Monte Carlo steps employed for the annealing
process between 10.000 K and 300 K.

\begin{center}
{\centering 
\begin{tabular}{|c|c|c|c|c|c|}
\hline
$MCS$ & 20 & 320 & 5.120 & 81.920 & 655.360 \\ \hline\hline
$\mu _{infl}$ & -3.14 & -3.12 & -3.11 & -3.11 & -3.11 \\ \hline
$\mu _{infl}^{ideal}$ & -3.12 & -3.11 & -3.11 & -3.11 & -3.10 \\ \hline
\end{tabular}
}
\end{center}

\textbf{Table 3} Chemical potential $\mu _{infl}$ in eV corresponding to the
isotherms resulting from the simulation of Ag deposition on kink sites.
These belong to Au imperfect surfaces generated by simulated annealing with
different number of Monte Carlos steps ($MCS)$. $\mu _{infl}^{ideal}$ are
the corresponding values for the inflection points of the ideal isotherms. %
\newpage\ 

\section{Figure Captions}

\textbf{Figure 1} a) Some of the most common surface defects found on single
crystal surfaces: 1) flat terrace, 2) monoatomic step, 3) kink site. b), c),
d) different steps of metal deposition.

\textbf{Figure 2} Environment of S sites ( a) S=9, b) S=13) surrounding an
adsorption node. These were employed for the calculation of the
adsorption/desorption energies of Ag and Au on Au(100) for a cutoff radii
corresponding to the distance between second (a) and third (b) nearest
neighbours. The particle adsorbed is located in the central box (9 and 13
respectively) .

\textbf{Figure 3} Adsorption energies of a Ag atom at the center of the
environment shown in figure 2b. The histogram drawn with the continuous line
considers only the configurations involving Ag atoms, while those
configurations involving only Au atoms were drawn with the dotted line. The
plot drawn with pulses includes all the possible configurations that may be
obtained with the environment including 13 adsorption nodes. In all cases
the proper normalizing factors were employed( 4096, 4096 and 531441
respectively).

\textbf{Figure 4} a) Adsorption isotherms for Ag deposition on defect free
Au(100) (1-4) and Ag(100) (5) single crystal surfaces. Different lattice
sizes $(N\times N)$ and cutoff radii $R$ were employed. In each case $R=n$
denotes that the distance to the $n^{th}$ nearest neighbour was employed. 1)$%
N=100$, $R=3$;  2) $N=40$, $R=2$;  3) $N=40$, $R=3$;  4) $N=100$, $R=2$;  5) 
$N=40$ and $R=3$.

\textbf{\ }b) Hysteresis effects for the simulation of Ag adsorption on
Au(100). The simulations were performed on a $100\times 100$ lattice. The
cutoff radius corresponded to $R=3$ and the starting conditions for the
coverage degree were 1)$\Theta _{Ag}=1$;  2) $\Theta _{Ag}=0.5$ and 3) $%
\Theta _{Ag}=0$

\textbf{Figure 5} Adsorption isotherms for Ag adsorption on Au(100) at
different temperatures. .

\textbf{Figure 6} Final configurations for the simulated annealing
simulations. The number of Monte Carlo steps $N_{MCS}$ increases from upper
left to down right. $N_{MCS}=20\times 2^{m-1}$, where $m$ is the ordinal
number of the configuration in the figure.

\textbf{Figure 7} a)Adsorption isotherms for Ag deposition on kink sites for
different types of imperfect surfaces obtained through the simulated
annealing procedure.

\textbf{\ }b)Ideal isotherms calculated according to the procedure indicated
in the text for the different types of imperfect surfaces.

\textbf{Figure 8} Types of kink sites most commonly found after the
simulated annealing procedure.

\textbf{Figure 9} Adsorption isotherms for decoration of monoatomic steps.
The Au surface structures considered were the final states of simulated
annealing runs with 20, 320, 5120, 81920 and 655360 $MCS$ per temperature
cycle respectively.

\textbf{Figure 10} Final state at different chemical potentials for a system
in which the Au islands were obtained by a simulated annealing process with $%
5120\ MCS $ per cooling cycle. The Au atoms on the surface are represented
by black pixels and the Ag atoms by grey pixels. 1) $\mu=-3.29\ eV$, 2) $\mu=-3.07\ eV$,
 3) $\mu=-2.97\ eV $, 4) $\mu=-2.94\ eV $, 5) $\mu=-2.93\ eV$, 6) $\mu=-2.91\ eV$

\textbf{Figure 11} Adsorption isotherms for Ag deposition on Au(100) in the
presence of surface defects. Each curve corresponds to one of the five
systems considered previously. \newpage\

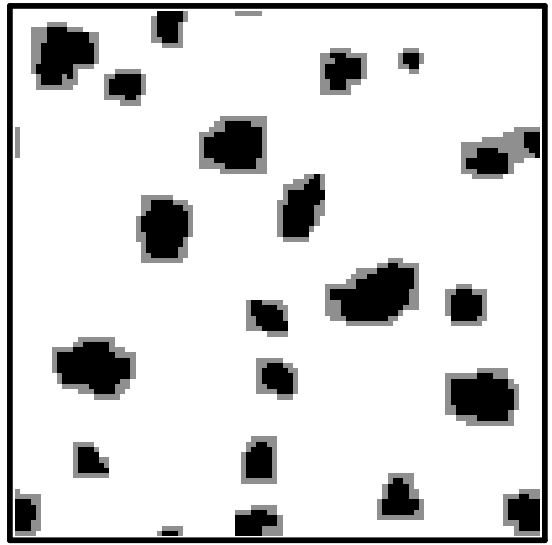
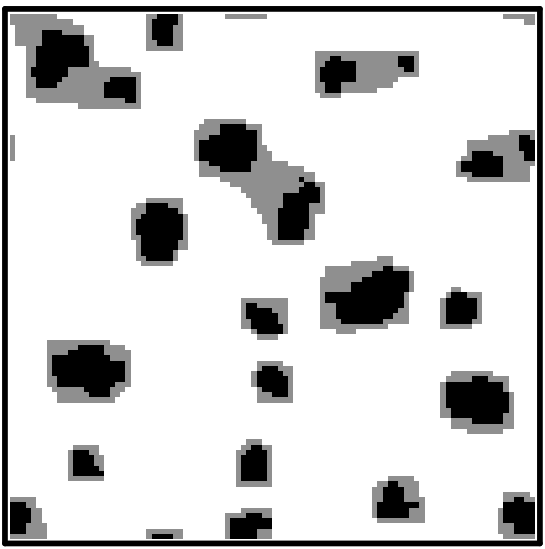
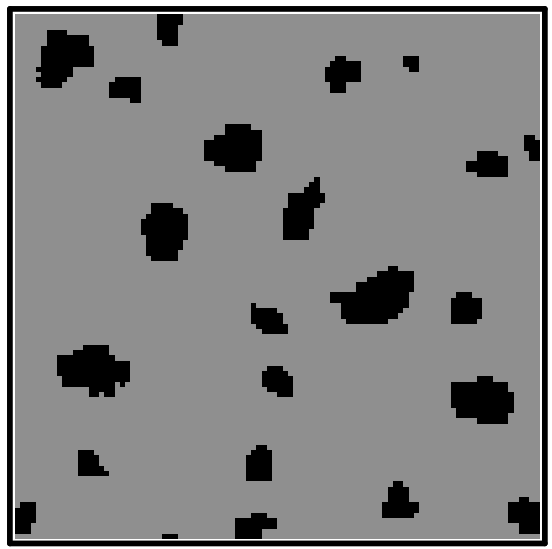
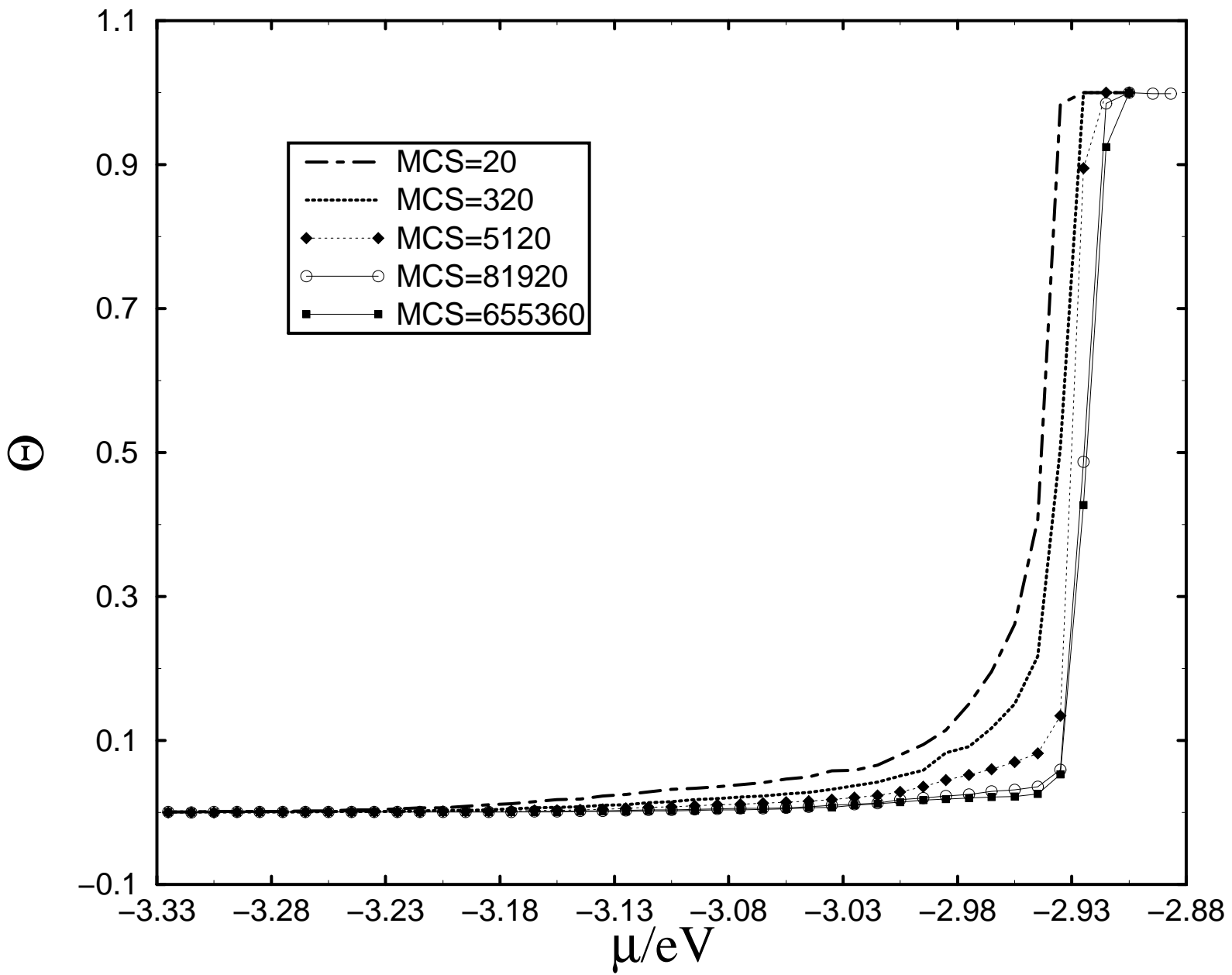
\end{document}